\documentclass[twocolumn,showpacs,preprintnumbers,amsmath,amssymb,prb]{revtex4}

\usepackage{graphicx}
\usepackage{dcolumn}
\usepackage{bm}

\newcommand{\etal}{{\em et al.}}

\begin{document}

\title{First-principles calculation of mechanical properties of Si $\langle$001$\rangle$ nanowires and comparison to nanomechanical theory}

\author{Byeongchan Lee}
\author{Robert E. Rudd}
 \email{robert.rudd@llnl.gov}
\affiliation{Lawrence Livermore National Laboratory, University of California, L-415, Livermore, California 94551 USA}

\date{\today}

\begin{abstract}
We report the results of first-principles density functional theory
calculations of the Young's
modulus and other mechanical properties of hydrogen-passivated
Si $\langle$001$\rangle$ nanowires.  The nanowires are taken to have predominantly
\{100\} surfaces, with small \{110\} facets according to the Wulff shape.
The Young's modulus, the equilibrium length and the constrained residual
stress of a series of prismatic beams of differing sizes are found to
have size dependences that scale like the surface area to volume
ratio for all but the smallest beam.  The results are compared with
a continuum model and the results of classical atomistic calculations
based on an empirical potential.  We attribute the size dependence to
specific physical structures and interactions.
In particular, the hydrogen interactions
on the surface and the charge density variations within the beam are
quantified and used both to parameterize the continuum
model and to account for the discrepancies between the two models and
the first-principles results.
\end{abstract}

\maketitle

\section{Introduction}
\label{sec:intro}

The mechanical response of structures at the nanoscale is known
to be different than that of their macroscopic analogs, and surface
effects in these high surface-to-volume devices are important.\cite{Cahn}
Significant strides forward have been made in the understanding of
these effects, but a predictive theory of nanomechanics remains
an open problem both at the academic level and in terms of
implementation in nanodevice design codes.
Nanoscale mechanical devices have been proposed
for applications for a range of nanoelectromechanical
systems (NEMS).  These devices include high-frequency
oscillators and filters,\cite{SiResonators1}
nanoscale surface probes,\cite{nanotubeProbe}
probes of single molecules \cite{singleMol}
and spins,\cite{singleSpin}
nanofluidic valves,\cite{WAG}
and q-bits for quantum computation.\cite{ClelandGeller, Blencowe}
The process of design and fabrication of these devices
is extremely challenging, requiring new techniques for
synthesis, characterization, integration and modeling
of the device performance that are currently the subject
of active research.  The process is complicated in part
by uncertainties about how even perfectly fabricated nanoscale
mechanical devices should behave due to our incomplete
theoretical understanding.

At sufficiently small sizes the fact that materials are comprised
of atoms and are not continuous media becomes important.
That size scale is quite close to atomic dimensions.
Somewhat larger nanoscale
structures may be described by continuum mechanics provided the
theory is suitably extended to account for the occurrence of effects
irrelevant in larger structures.\cite{Cahn} One class of these
effects is related to surfaces.  Since the surface area to
volume ratio grows as the size of a structure is decreased,
surfaces are expected to play a more prominent role at the
nanoscale.

Of the various mechanical properties, the Young's modulus is of particular
interest
as an important parameter in the function of nanoscale devices
such as flexural-mode mechanical resonators \cite{SiResonators1}
and as an archetype in terms of the scaling behavior
of a variety of mechanical properties including the Poisson ratio,
the anelastic damping coefficients and so on.  The Young's modulus
is defined as the ratio of the stress applied to stretch a cylindrical
(or prismatic) beam to the resulting elongation strain.  For bulk
materials it may be expressed in terms of the bulk elastic constants
$C_{ijkl}$ and it is a material constant, independent of the size
of the structure.  If the system is comparable in size to any mechanical
inhomogeneities it contains, such as grains or inclusions,
the the modulus may exhibit a size dependence; interestingly, at the
nanoscale even a system free from internal inhomogeneities has been
predicted to have a size-dependent Young's modulus.\cite{BMVK,quartzResonators}

Here we present in detail the results of an {\em ab-initio} study of the
mechanical properties of silicon nanowires.  Our goal is to
calculate these properties from first principles, using a quantum-mechanical
description of the electronic binding that is free from any empirical
input or other {\em a priori} assumptions about the nature of the bonds.
We then compare the results to existing nanomechanical models of the
size dependence of the properties of nanosystems.  This comparison
requires the calculation of structural and mechanical properties of
various reference systems, which we also report here.
Some of the nanowire results have
been presented previously in a more concise form.\cite{LeeRudd}

While our principal focus is on the Young's modulus,
we also calculate and analyze the residual stress
and equilibrium elongation of the nanowires.
We consider prismatic Si [001] nanowires with a combination of
$\{$100$\}$ and $\{$110$\}$ hydrogen-passivated surfaces, and
single crystal cores as in
experiment.\cite{SiResonators1,SiResonators2} We have
chosen the [001] orientation for the longitudinal axis because
of its relevance to the NEMS devices;\cite{SiResonators1}
Si nanowires grown rather
than etched typically have different orientations.\cite{Lieber}
Hydrogen passivation results from rinsing the oxidized
Si surfaces with HF. It provides a standard system suitable for
a systematic study of the size dependence in nanomechanics.
With other surface conditions the band gap can vary greatly, and nanowires
can go from semiconducting to metallic;\cite{metallicSiNW}
whereas the H-passivated wires remain
semiconducting \cite{SiNWSurface} down to the smallest sizes studied
here and the surfaces do
not change the nature of
Si-Si chemical bonding from its covalent character.

The article is organized as follows.
We begin with a review in Section \ref{sec:nanomech}
of the state of our current understanding
of the mechanics of nanoscale structures focusing on
nanowires, especially nanowires composed of Group IV elements.
Then we report in Section \ref{sec:ref} the
result of calculations of Si bulk and surface systems that are
used later in the analysis of Si nanowire mechanical properties.
In Section \ref{sec:geometry} we describe the geometry and
configuration of the nanowires studied here.  We describe how
the series of wires of various sizes was created, and note
subtleties in the description of the wires with continuum quantities.
In Section \ref{sec:nwcals} we present and analyze the
residual stress of the nanowires
constrained from longitudinal relaxation.  We analyze this
stress and the related property of the equilibrium length of
the wire.  Finally,
we report the result of calculation of the Young's
modulus, and analyze the size dependence. The basic calculations of the
Young's modulus, the residual stress and the equilibrium length
were reported briefly in Ref.\ \onlinecite{LeeRudd}. Here we present
the results for the reference systems and the detailed analysis
of the mechanics not included in the short article.  This analysis
goes beyond the successful comparison of the first-principles results
with the scaling laws derived from continuum mechanics models including
surface effects presented previously\cite{LeeRudd}
and allows the determination of the specific physical
structures and interactions at the atomic and electronic level that lead
to the size dependence.  Many, but not all, of these structures and
interactions are present in the surfaces of the slab reference systems
that may be analyzed at much less computational cost than the nanowires.
The first-principles calculations allow us to assess in detail the
validity and point of breakdown of the continuum models and the physics
captured therein.

\section{A brief review of nanomechanics}
\label{sec:nanomech}

The mechanical properties of nanowires are expected to depend
on the size (diameter) of the wire
because surface effects increasingly dominate as
the devices are miniaturized down to the nanoscale.
This expectation has been borne out by computer simulation
using atomistic techniques based on empirical potentials.
The first such calculations were done for
single-crystal $\alpha$-quartz beams, finding a systematic
size dependence in which the Young's modulus
decreased with decreasing size.\cite{BMVK,quartzResonators}
These and calculations of the Young's modulus for various
other materials have predicted a size-dependent modulus
with an additive correction to the bulk value that
scales like the surface area to volume ratio.\cite{Miller,RuddIJMSE}

Continuum-based formalisms for nanoscale mechanics have been proposed
that include the effect of surface properties on the mechanical behavior.
One class of models is based on the surface free energy and its
first two strain derivatives (the surface stress and the surface
elastic constants).\cite{Gurtin,Miller,Kukta}
Other approaches make use of the Cauchy-Born rule
applied to surfaces to
avoid the need for precalculated surface properties for
metals\cite{surfaceCauchyBorn} and a Kirchoff rod model for the
properties of torqued amorphous nanowires.\cite{Fonseca}
Beyond the inclusion of surface properties,
there have been efforts to explore the relevance of
other sources of size-dependence.
A few studies claim an additional contribution that scales like the edge
to volume ratio (cf.\ Ref.\ \onlinecite{BMVK}):
 such a contribution, with
a factor of the logarithm of the separation of
the edges, has been discussed for
epitaxial quantum dots.\cite{Shchukin,RuddQDot,RuddQDot2}
Another study proposes the size dependence of the Young's
modulus due to the anharmonicity (nonlinearity) of the
bulk elastic moduli together with the strain resulting from the
surface stress.\cite{bulkAnharmonicity}

An intuitive way of understanding these effects
is that there is a layer of material at the surface (and edges)
whose mechanical properties differ from those of the bulk
including different elastic moduli and eigenstrains.
The term eigenstrain here means that the surface layer is
constrained by its interface to the bulk to be at a nonzero
strain with respect to its minimal energy state; i.e., at
a nonequilibrium lattice constant if the surface layer
is crystalline.  The surface layer could be chemically
distinct from the bulk, such as an oxide layer or a
hydrogen-passivated surface, but the effect may be entirely due to
the structural difference at the surface, such as a bare
reconstructed surface. These surface effects could possibly
force the system to deviate significantly from the bulk
equilibrium, and go out of the linear elastic regime.
Atomistic calculations provide insight into these effects,
but then we must assess to what extent the results from
empirical potentials can be trusted.  For many multicomponent
systems, empirical potentials are either not available or
not validated.  Even for pure Si nanowires, there is physics
missing from empirical potentials that may be crucial, possibly
to be included in the future in generalized potentials such as done
recently for platinum nanoparticles.\cite{LeeSS} One
example is the buckled dimer reconstruction on the Si $\{$100$\}$
surface that is not the ground state of any of the standard
Si empirical potentials,\cite{Balamane} or even a tight-binding
model.\cite{Arias98}

To date, experimental data on the size dependence of nanostructure
mechanics are very limited.  If fabricating the nanoscale structures
and measuring their mechanical properties such as the Young's modulus
is difficult, then the difficulty is compounded in obtaining
reproducible measurements free from systematic error across a
series of structures of decreasing size.  Promising work has
begun in this direction.
Atomic force microscopy (AFM) measurements
of the Young's modulus\cite{LieberMechanics} ($E$)
of cast metallic nanowires with diameters in the range of 30 to 250 nm
show a strong size dependence.\cite{Cuenot}
Recent experiments have also found a strong size dependence for
$E$ of ZnO nanowires with diameters
in the range of 17 to 550 nm.\cite{ZnOnanowire}

For semiconductor wires,
Measurements of the Young's modulus and the bending modulus of
crystalline boron nanowires with diameters of 40--58 nm and
43--95 nm, respectively, have shown no systematic size
dependence.\cite{RuoffB}
The Young's modulus of single crystal germanium nanowires
with the diameters of 40--160 nm
is also found to be comparable to the bulk value.\cite{GeWires}
A study using a different AFM technique reported a value of $E$
of $18\pm2$ GPa for irregularly shaped $<10$-nm Si $\langle$100$\rangle$
nanowires;\cite{Kizuka}
for Si$\langle$111$\rangle$ wires with 100--200 nm in diameter,
$E$ has been found to be consistent
with the bulk value;\cite{PYangSiNW} Si $\langle$110$\rangle$ nanowires
with diameters of 12--170 nm show size-dependent softening.\cite{Li110NW}
In another Group-IV system, measurements of $E$ for silica
nanobeams have demonstrated that the way in which the beam is clamped
(i.e., the boundary conditions) affects the apparent value.\cite{RuoffSiO2}
Experimental challenges measuring the intrinsic nanoscale Young's modulus
make this a topic of continued activity, leveraging earlier work on the
mechanics of nanotubes.\cite{nanotubeMech}

In the absence of definitive experimental data, first-principles quantum
mechanical calculations based on density functional theory (DFT)
can provide a robust prediction for the
behavior of the nanoscale structures, but there have been
few results reported.  True first-principles techniques do not
rely on any empirical data, solving the quantum-mechanical Kohn-Sham
equations of DFT to achieve predictions from first principles.\cite{KohnSham}
One quantum study based on an empirical
tight-binding technique has been published.\cite{Arias98,Arias}
Recently the size dependence of the Young's modulus of thin slabs
has been reported calculated from first-principles,\cite{slabs,Gumbsch}
results that are quite relevant but not equivalent to nanowire
calculations due to the absence of edges.
We are not aware of any {\em ab initio} calculations of nanowire
moduli in the literature apart from the brief article,
Ref.~\onlinecite{LeeRudd}.

\section{Bulk and surface reference calculations}
\label{sec:ref}

\begin{table}
\caption{\label{tab:bulkProperties}Bulk elastic properties calculated
with DFT. The Young's modulus ($E$) and the Poisson's ratio ($\nu$)
have been derived from the elastic moduli $C_{11}$ and $C_{12}$.
All units are in GPa except for Poisson's ratio,
which is dimensionless.}
\begin{ruledtabular}
\begin{tabular}{cccc}
& This work & Other theory\cite{Gumbsch} & Experiments\cite{elasticConstants}\\
\hline
$C_{11}$& 154.6 & 154 & 167.7\\
$C_{12}$& 58.1 & 55 & 65.0\\
$C_{44}$& 74.4 & - & 80.4\\
$E_{bulk}$ & 122.8 & 125.1 & 131.4 \\
$\nu_{bulk}$ & 0.27 & 0.26 & 0.28 \\
\end{tabular}
\end{ruledtabular}
\end{table}

Nanowires have structural aspects ranging from bulk-like atomic
arrangement in the core of wires to the more open, and often
significantly relaxed, surface and edge structures.  As the
size of the wire decreases, the surfaces and edges play an
increasing role.
In this section we establish the reference properties of the bulk
silicon and the surfaces needed to analyze the mechanics of
hydrogen-passivated Si $\langle$001$\rangle$ wires.
The reference data will help us
understand the complex mechanics of nanowires in terms of simpler physics
and assess how continuum surface physics parameterized by the properties
of the reference systems can augment bulk continuum mechanics
to provide a robust description of nanometer-scale structures.

\subsection{Bulk properties}
\label{subsec:bulk}

A series of calculations has been done to obtain the
bulk properties of crystalline silicon.
The history of density functional theory investigations of
silicon is long, going as far back as the
late 1970's.\cite{bulkSi1}  The elastic constants of silicon
were calculated from first principles in the
work by Nielsen and Martin,\cite{bulkSi2} and anharmonic
effects in silicon have been mentioned in the
early work by Ihm and coworkers.\cite{bulkSi3}  Our purpose in
performing similar calculations in the context of this study
is to provide an assessment of accuracy of our results compared
to literature values, and to provide reference numbers calculated
using the same code and techniques as in the nanowire calculations
in order to provide directly comparable numbers to analyze
the nanowire results.

We use first-principles density functional theory (DFT):
specifically, the Vienna ab-initio simulation package (VASP)
using the projector augmented-wave method (PAW) \cite{PAW1,PAW2}
within the generalized gradient approximation (GGA) by
Perdew and coworkers.\cite{GGA}
The PAW potentials with 4 valence electrons (3s$^2$3p$^2$) are used.
The energy cutoff for the plane-wave expansion is 29.34~Ry, and
12$\times$12$\times$12 Monkhorst-Pack mesh \cite{kPointSampling}
is used for $k$-point sampling. The system consists of the 8 atoms of
a single Si diamond cubic unit cell with periodic boundary conditions.
The supercell is deformed to the appropriate strain and the
atomic positions are fully relaxed for each calculation.

From the bulk crystal under uniaxial loading, i.e., stressed
along the $[$001$]$ direction and stress-free in two transverse directions,
the Young's modulus and Poisson's ratio of the bulk reference system
have been obtained. The equilibrium Young's modulus
calculated from derivatives of a fit of the total energy in
the strain range from 5\% compression to 5\% tension is 122.5~GPa
using a fifth-order polynomial fit,
the corresponding Poisson's ratio is 0.27. The value does not change
significantly for a lower order fit. The difference between the
second order fit and the fifth order fit is less than 1\%. The
negligible modulus difference between harmonic approximation and
anharmonic expansion implies that the effect of bulk anharmonicity on
the Young's modulus is small at these strains.
Another way to assess the effect of anharmonicity is to examine
the compression-tension asymmetry of the modulus using the
fifth order fit: 5\% tension in the $[$001$]$ direction results in
mere 3\% anharmonic increase in the Young's modulus.
It is well known that the covalently bonded silicon has a
weak anharmonicity compared to metals, as typically expressed
in terms of a relatively small Gr\"{u}neisen parameter.\cite{Gauster}
We show below based on these reference calculations that the
anharmonicity does not play an important role in the size
dependence of the Young's modulus for silicon nanowires,
but the conclusion might have been different in a more strongly
anharmonic system.\cite{bulkAnharmonicity}

We have also computed $C_{11}$, $C_{12}$ and $C_{44}$ from separate
calculations.  The values we obtain are $C_{11}$=154.6 GPa,
$C_{12}$=58.1 GPa and $C_{44}$=74.4 GPa,
in good agreement with the previous
DFT calculations, and are $\sim$10\% less than the corresponding experimental
values.\cite{elasticConstants} The Young's modulus and Poisson's ratio have
been alternatively obtained from $C_{11}$ and $C_{12}$, and listed in
Table~\ref{tab:bulkProperties}. The direct numbers
from strictly uniaxial stress, and indirect numbers from cubic elastic
constants are essentially identical.

\subsection{\label{sec:surface}Surface properties}
\label{subsec:surface}

\begin{table*}
\caption{\label{tab:surfaceProperties}Surface energies, stresses, and
elastic constants for Si \{100\} surface with symmetric and canted
dihydride phases, and
Si \{110\} surface calculated with DFT. For surface elastic constants,
[001] is taken to be
the principal direction for $S_{11}$, and the other in-plane
direction orthogonal to [001]
is taken to be the second direction for $S_{22}$. The surface
modulus, $S$, is analogous to the bulk modulus,
and defined as $S = (S_{11}+S_{22}+2S_{12})/4$ when isotropic strain
is applied, i.e., $\epsilon_{11}=\epsilon_{22}$.}
\begin{ruledtabular}
\begin{tabular}{ccccccc}
& surface energy & surface stress & \multicolumn{4}{c}{surface elastic constants (eV/\AA$^{2}$)}\\
& (meV/\AA$^{2}$) & (meV/\AA$^{2}$) & $S_{11}$ & $S_{22}$ & $S_{12}$ & $S$\\
\hline
\{100\} symmetric & 28.5 & -123.2 & -1.191 & -1.191 & 1.919 & 0.364\\
\{100\} canted & 19.6 & -55.0 & -0.659 & -0.659 & 0.457 & -0.101 \\
\{110\} & 8.2 & -1.3 & -1.223 & 0.354 & -0.614 & -0.526 \\
\end{tabular}
\end{ruledtabular}
\end{table*}

\begin{figure}
\includegraphics[width=0.25\textwidth]{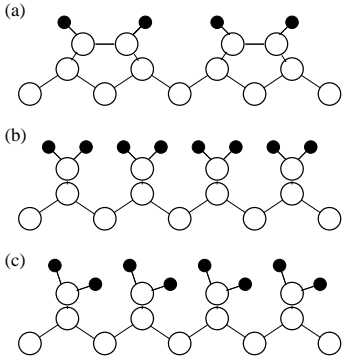}
\caption{\label{fig:passivation}Hydrogen passivation on Si(100) surface.
(a) 2$\times$1 monohydride, (b) 1$\times$1 symmetric dihydride, and
(c) 1$\times$1 canted dihydride.}
\end{figure}

\begin{figure}
\includegraphics[width=0.45\textwidth]{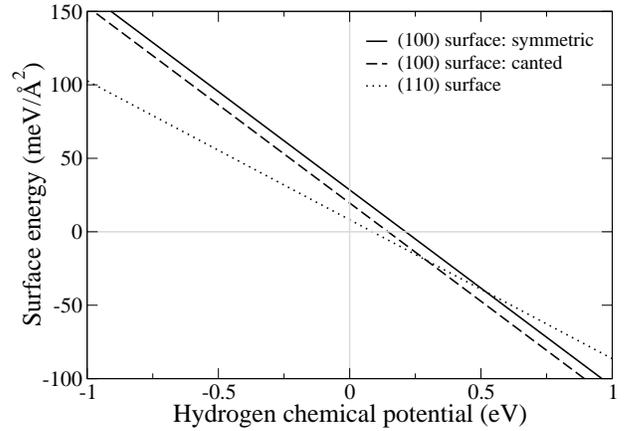}
\caption{\label{fig:surfaceEnergy}
Surface energies of the symmetric
and canted H-terminated Si(100) surface
and the H-terminated Si(110) surface
as a function of hydrogen chemical potential as calculated in DFT. The
hydrogen chemical potential is calculated relative to molecular hydrogen
dissociation energy.}
\end{figure}

We now turn to the properties of the H-terminated Si surfaces so
important to the nanowire mechanics. The ground states of low-index Si surfaces
have long been studied both for bare (clean) reconstructed
surfaces\cite{bareSurface1,bareSurface2,bareSurface3} and for
hydrogen-passivated surfaces.\cite{Sakurai,cantedSurface} In recent
years, after the ground states had been identified, the attention has
switched to the mechanical properties of the bare surfaces for
the application to Si nanowires using empirical potentials,\cite{Miller}
and to Si thin slabs using first principles calculations.\cite{Gumbsch}
Despite considerable attention to Si surfaces, the mechanics of
hydrogen-passivated surfaces has not been studied in detail.

The mechanics of the hydrogen-passivated surfaces, together with the mechanics
of the bare surfaces, give an indication of the range in which the surface
mechanics in the real world resides.  The two surface conditions represent
two opposing ideals, i.e., perfect passivation of very reactive bonds and
no passivation of such dangling bonds. This idealization motivates the
need to understand the mechanics of the hydrogen-passivated surfaces
qualitatively as well as quantitatively as a basis for further study
of the mechanics of nanowires, not to mention many other NEMS devices.
In principle, the hydrogen-passivated surfaces still have a surface effect,
similar to those from the bare wires, that arises from having a free surface.
This is an intrinsic effect. In addition, they have another effect due to
the surface hydrogen atoms, an extrinsic effect. The surfactant-induced
stress has been explained within the context of local electronic environment
for semiconductor surfaces,\cite{Vanderbilt,Ibach} but it is primarily due
to the interaction between adsorbate and substrate atoms. Hence, it is less
important for hydrogenated Si surfaces, where the hydrogen-hydrogen
interaction is prominent as evidenced in the surface (ground-state)
structures.\cite{cantedSurface,SiSlabs}
The efforts to understand the surface mechanics using
experimental\cite{experIndent} and simulated\cite{simulIndent}
nanoindentation provide qualitative descriptions but the accurate
(size-dependent) Young's modulus or the surface elastic constants have
proved difficult to access via this indirect technique, leading to a
need for direct measurement. In the work presented here first-principles
calculations are crucial to deal properly with quantum mechanical effects
such as the hydrogen-hydrogen interaction, and to investigate directly the
size-dependent mechanical properties of the hydrogenated surface in
a quantitative sense.

Thus, we investigate on the surface properties of the low-index
facets relevant to $\langle$001$\rangle$ nanowires,
Si $\{$100$\}$ and $\{$110$\}$ surfaces. A few surface
patterns are known for the hydrogen-passivated $\{$100$\}$
surface\cite{cantedSurface} and illustrated in Fig.~\ref{fig:passivation};
we have focused on the 1$\times$1
dihydride phase partially because some of the wires in the range
of interest may be too small to develop any larger pattern fully, and
partially because our focus in this work is on size effects, and thus
we must have the same pattern for all sizes. It is also known that
the dihydride phase is preferred over the monohydride phase
at low temperature.\cite{dihydrideCiraci}

The surface energies of hydrogenated Si surfaces
we calculated in DFT are shown in Fig.~\ref{fig:surfaceEnergy},
with the three curves corresponding to the $\{$100$\}$ symmetric
dihydride surface, the $\{$100$\}$ canted dihydride surface,
and the $\{$110$\}$ surface.
Again VASP was used, with 29.34 Ry cutoff energy and
12$\times$12$\times$1 Monkhorst-Pack mesh in the two-dimensional Brillouin zone.
The calculations were performed on slabs with 14 layers of Si for
the $\{$100$\}$ slabs and 15 layers for the $\{$110$\}$ slabs
with a single unit cell in plane,
for a total of 14 Si and 4 H atoms, and 30 Si and 4 H atoms, respectively.
The atomic positions were relaxed until
the force on each atom was less than $10^{-2}$ eV/{\AA}.

The surface energies depend on the chemical
potential of hydrogen,\cite{cantedSurface}
and the hydrogen chemical potential is taken relative to
the hydrogen molecule dissociation energy; i.e., zero chemical potential
implies that the hydrogens dissociate from the molecule and
attach to the surface to terminate the dangling bonds without any
cost in energy. The zero-point-energy term\cite{cantedSurface} is omitted, but
this term only shifts the energy and does not alter the physics. The
choice of the reference chemical potential
is somewhat arbitrary but this choice measuring the chemical potential
relative to the H$_2$ dissociation energy
facilitates comparison of the two
$\{$100$\}$ surfaces: a change in the hydrogen
chemical potential adds the same offset to both surface energies, and
only the surface energy difference matters. The horizontal line
represents zero surface energy, and below this line the surface
energy is negative; i.e., surface area is maximized at the cost
of material cohesion. We are interested in the region above the
zero-surface-energy line, and we find that the $\{$110$\}$ surface is
preferred over the other two surfaces.

The surface stress can be unambiguously
determined since it is a strain
derivative of the surface energy, and hence the chemical potential dependence
goes away. The surface stress has been calculated for various materials
previously using first-principles\cite{Needs}
and classical atomistic\cite{Streitz} techniques.
As can be seen in Table~\ref{tab:surfaceProperties}, the
surface stress is twice larger for the symmetric $\{$100$\}$ surface
stress than for the canted $\{$100$\}$ surface. The strong H-H repulsion
between neighboring hydrogen atoms is relaxed by tilting the dihydride,
and the surface stress is reduced. The negative stress indicates compression.
The surface stress of the $\{$110$\}$ surface is essentially negligible.
Across a range of loading conditions and fitting procedures, its values
are small, varying little
and what variation there is most likely comes from a numerical artifact:
the surface stress under $[$001$]$ strain ranges
from -1.3 to -3.2 meV/\AA$^2$ depending on the order of fitting.
The resulting uncertainty due to the fitting in the equilibrium surface lattice
spacing is less than $10^{-3}$ {\AA} for the 15-layer $\{$110$\}$ slab.

The next derivative (second strain derivative) of the surface energy
is the surface elastic constant. Similar to bulk elastic constants, surface
elastic constants account for the material stiffness. Three surface
elastic constants,
$S_{11}$, $S_{22}$ and $S_{12}$ have been calculated with the principal
direction of $[$001$]$. The negative constants indicate softening due
to the surface, and the opposite signs of $S_{11}$ and $S_{12}$ can
be thought of as
a negative Poisson effect. For example, a higher positive $S_{12}$ for
the symmetric $\{$100$\}$ surface means that, when the surface slab is
stretched in the $[$001$]$ direction, the slab would expand
(direction of energy reduction) in the transverse direction
as well due to the particular alignment of the
dihydride in the [110] direction. On the other hand, a positive
Poisson effect has been observed for the $\{$110$\}$ slab. This conventional
result is expected as there is virtually no H-H repulsion
on that facet.

Also presented is the surface modulus, $S$, a two dimensional counterpart
of the bulk modulus under hydrostatic loading. As expected, the
$\{$100$\}$ symmetric surface has the highest surface modulus,
the $\{$100$\}$ canted surface is next, and the $\{$110$\}$ surface
shows the lowest modulus: the hydrogen repulsion is highest for the
$\{$100$\}$ symmetric surface, and it is virtually zero for the
$\{$110$\}$ surface. The strong repulsion of the symmetric dihydride
even leads to a positive surface modulus, implying that, in the case
of isotropic plane stress
the $\{$100$\}$ symmetric
surface slab is even stiffer than the bulk system with the same
volume. However, the increase in individual elastic constants
shows a different tendency than that of the surface moduli: the $S_{11}$
is highest for the $\{$100$\}$ canted surface.
We believe that this may be explained by the relaxation of the hydrogen
repulsion during uniaxial strain.
The uniaxial strain in the [001] direction induces the shear
strain for the $\{$100$\}$ surface unit cell, and dihydrides
become misaligned and consequently the associated repulsion is
more or less relaxed.
In this situation, the H-H repulsion
may be substantially relaxed with respect to the applied strain for the
$\{$100$\}$ canted surface. On the other hand, $\{$100$\}$ symmetric
surface seems to undergo a smaller change in the H-H interaction
energy and its contribution to $S_{11}$ is also smaller.  Again, this
change
has nothing to do with the absolute magnitude of hydrogen interaction.
Rather, the magnitude of modulus is directly related to the interaction
energy change with respect to the given strain, in particular
the curvature with respect to the applied strain.

To summarize, the stiffening effect of the surface hydrogen for
the $\{$100$\}$ canted surface is equally noticeable under uniaxial and
biaxial strain, but the effect seems prominent under biaxial strain for
the $\{$100$\}$ symmetric surface. In other words, an arbitrary strain
would relieve the hydrogen repulsion for the $\{$100$\}$ canted surface,
but a uniform biaxial strain would dominantly do it for the $\{$100$\}$
symmetric surface. This complexity of the hydrogenated $\{$100$\}$ surface
comes from the
fact that the principal crystallographic directions of the bulk are
$\langle$001$\rangle$ directions, whereas those of the $\{$100$\}$ dihydride surface
are $\langle$110$\rangle$ directions: the principal directions from
one perspective are the maximum shear directions from the other.

The two hydrogenated $\{$100$\}$ surfaces, symmetric and canted,
have substantial differences in their surface stress and surface moduli,
and the resulting equilibrium surface lattice spacings for the 14-layer
slab ($\sim$1.96 nm thick) are 1.83\% and 0.76\% longer than the
bulk lattice, respectively. The 15-layer $\{$110$\}$ slab
($\sim$2.95 nm thick) exhibits the elongation on the order of
0.01\%, i.e., less than $10^{-3}$ {\AA} difference from the bulk spacing,
due to its negligible intrinsic surface stress.

\subsection{Silane chains}
\label{subsec:silane}

\begin{figure}
\includegraphics[width=0.45\textwidth]{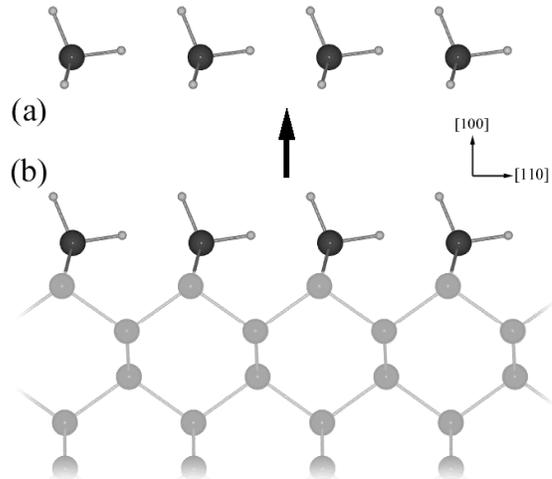}
\caption{\label{fig:silaneChain} (Color online)
One dimensional periodic chain of silanes
and its relationship to the (100) surface.
The silane chain, (a), has been taken from the fully-relaxed hydrogenated (100) surface
with canted dihydrides, (b), and the backbonds have been passivated with
additional hydrogens.}
\end{figure}

\begin{figure}
\includegraphics[width=0.45\textwidth]{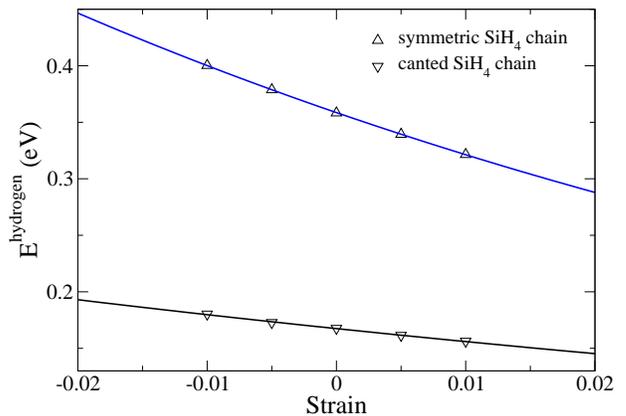}
\caption{\label{fig:extrinsic} (Color online)
SiH$_4$-SiH$_4$ interaction energy
calculated in DFT as a function of strain
for the symmetric and canted chains. The solid curve is an exponential fit to
the presented data.}
\end{figure}

In order to separate the hydrogen interatomic (H-H) interaction contribution
from the other contributions to the surface elastic constants,
the effect of surface hydrogen is evaluated from a periodic chain of
silane molecules, in which the dominant H-H repulsive interaction is
coming from neighboring silane molecules. The periodic chain is taken
from the fully relaxed $\{$100$\}$ surface slab as illustrated in
Fig.\ \ref{fig:silaneChain}, and the backbonds are terminated with 2
additional hydrogen atoms. The bond angles of the relaxed slab geometry
are preserved, and the backbond length is chosen to be
1.49~{\AA} based on the structure of an isolated SiH$_4$ molecule.
The canted silane chain is shown, but the same
procedure is applied to the symmetric chain.
The extrinsic contribution to the surface elastic constants from the H-H
interaction is given approximately by
\begin{equation}
S_{11}^{H-H} = \left. \frac{1-\nu}{\sqrt{2}A}\partial ^2
U^{H-H}/\partial \epsilon _{zz}^2\right|_{0}
\label{eq:extrinsicModulus}
\end{equation}
where $U^{H-H}$ is the H-H contribution to the total energy
and $A$ the area covered by one silane.
The prefactor is due to the decomposition into the longitudinal
and transverse directions of the wire including the
Poisson effect with $\nu \thickapprox \nu_{bulk} = 0.27$.
$U^{H-H}$ can be approximated by the
silane-silane interaction energy in the chain and given as
\begin{equation}
U^{H-H}(\epsilon) \thickapprox  U^{chain}[N, \epsilon]/N - U^{silane}
\label{eq:extrinsicEnergy}
\end{equation}
where $U^{chain}[N]$ is the total energy of the $N$-molecule chain of SiH$_4$,
and $U^{silane}$ is the energy of a single SiH$_4$ molecule.
$U^{H-H}$ can be interpreted as a energy penalty
to bring repulsive
silane molecules into a single chain. It is a reasonable
approximation in the sense that
the underlying physics is essentially identical in both cases:
H-H exchange repulsion is the dominant interaction, and both the
SiH$_2$ group on the $\{$100$\}$ surface and SiH$_4$ in the silane
chain have almost identical local environments.

$U^{H-H}$ as a function of strain is obtained from a series of
silane chains, whose geometries are taken from the corresponding
relaxed $\{$100$\}$ surface slabs. Due
to the nature of H-H repulsion, $U^{H-H}$ is fitted to an
exponential function around the bulk lattice spacing as shown in
Fig.\ \ref{fig:extrinsic}. The fit of an exponential form to the
data yielded the following dependence:
\begin{eqnarray}
U^{H-H}_{{\mathrm{symmetric}}} & = & 0.359~\exp(-10.97\epsilon)\\
U^{H-H}_{{\mathrm{canted}}} & = & 0.167~\exp(-7.10\epsilon)
\end{eqnarray}
in units of eV.
The ideal $S_{11}^{H-H}$ obtained from
the silane chains using Eq.~\ref{eq:extrinsicModulus}
for the symmetric $\{$100$\}$ and canted $\{$100$\}$
slabs are 1.488 eV/\AA$^2$ and 0.291 eV/\AA$^2$, respectively.

\section{Geometry of nanowires}
\label{sec:geometry}

The cross-sectional shape of the Si $\langle$001$\rangle$ wires we study
is a truncated square
with four $\{$100$\}$ facets and four $\{$110$\}$ facets. Some
wires studied here have no $\{$100$\}$ facets; for those that do, the ratio
of the facet areas is taken to be roughly in accordance with the
Wulff shape for a bare wire with $\{$100$\}$-p(2$\times$2)
and $\{$110$\}$-(1$\times$1) surface
reconstructions; i.e., the ratio of $\{$100$\}$ to $\{$110$\}$ area is 3.5:1.

The reason for using the Wulff shape of bare nanowires is that the effect
of surface condition on the material stiffness can be addressed by comparing
hydrogen passivated wires and bare wires. In the Wulff shape of
hydrogen-passivated wires, the $\{$110$\}$ area would be larger than
the $\{$100$\}$ area due to the lower energy of that facet, opposite
to the Wulff shape of bare nanowires used here.  It is advantageous to use
the bare-wire shape not only because the surface energy of a
bare slab can be unambiguously determined whereas that of a slab with
surface adsorbates such as hydrogen depends on the adsorbate chemical
potential, but because having a common cross section allows us to
focus on the size effect or the role of surface condition apart from
any effect due to shape.
The detailed comparison of the bare and H-terminated wires
will be given elsewhere.\cite{BCLeeInPrep}

\begin{figure}
\includegraphics[width=0.45\textwidth]{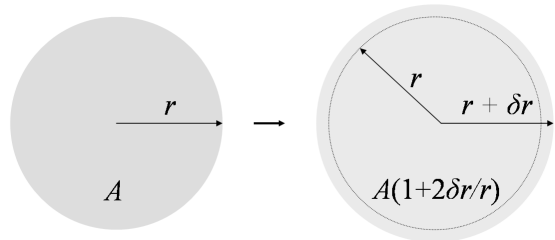}
\caption{\label{fig:crossSection}Size effect of the definition of
the cross-sectional area. The dotted circle in the
second circle indicates the initial size whose cross-sectional area is $A$.}
\end{figure}

The electronic structure of nanowires is one dimensional.
It requires a large electronic excitation to probe the transverse dimensions
of the wire.  The mechanics of the nanowire, while not fully
three dimensional, has unavoidable three dimensional aspects to
it.
The nanowires have a finite cross-sectional area that enters into
many of their mechanical properties and because of this cross section
they are able to support mechanical bending moments.
The relevance of the transverse dimensions to mechanical properties
poses a challenge.  The mechanical properties are most succinctly
phrased in terms of continuum mechanics, but this requires the
definition of continuum measures such as the cross-sectional area and the
transverse dimensions in terms of atomistic properties such as
ionic positions or electron densities.

To calculate the Young's modulus, for example, {\em we define the
cross-sectional area to be the area
bounded by the centers of the outermost (H) atoms}.
This choice is motivated by the fact that the volume excluded
by the beam from access by outside atoms is determined from
the forces arising from electron interactions. In other words,
any experimental measure of volume is based on the probe being strongly
repelled
due to electron interactions rather than between those of nuclei.
Therefore, a system that can be thought of as a discrete system
in an atomic description is continuous in an electronic description.
It is unusual, however, to apply the techniques of continuum mechanics
at subatomic length scales,\cite{Pask} and indeed
conventional scale-free continuum mechanics breaks down at nanometer scales
or higher because of the lack of physics relevant to small structures
such as surface energies and surface stresses.

The surface definition used here ensures that
most of the electron density is enclosed by the surface, the
boundary formed by H atoms, and the electron density from
Si atoms essentially vanishes beyond this point. In addition, the
positions of the nuclei are well defined and not subjective.
Other definitions of the bounding surface exist: for example,
the midplane between two identical H-passivated surfaces at
their minimum energy separation.\cite{SiSlabs}

Had we taken a different definition of
the cross-sectional area, the apparent size dependence of $E$
would have been different,  an ambiguity associated
with describing discrete atomic systems with continuum mechanics.
To illustrate the issue, consider the
effect on the size dependence of $E$
from changing the definition of cross-sectional area
by modifying the position of the surface $r$ by $\delta r$ as shown in Fig.\ \ref{fig:crossSection};
e.g.\ $\delta r$ would be the atomic radius for
a surface going through atomic centers vs.\ one circumscribing
their electron cloud.  The change $\delta r$
takes $A$ to $A'\approx A(1+2\delta r/r)$
\begin{eqnarray}
E(r)   & = & \frac{F}{A \epsilon} \\
E'(r)  & = & \frac{F}{A' \epsilon}
   \\ & \approx & \frac{F}{A\left( 1 + 2\delta r / r\right) \epsilon}
   \\ & = & E \left( 1 - 2\delta r / r + \cdots \right)
\end{eqnarray}
where $F$ is the force carried by the beam and $\epsilon$ is the
resulting strain.  The leading change to the Young's modulus,
$-2E(\infty)\,\delta r/r$, is proportional to the surface area to volume
ratio. This change modifies the value coefficient of the
surface-area-to-volume ratio term in the Young's modulus, so when
we calculate this coefficient, it is implicit that its value
is with respect to our prescription for the location of the surface.
The uncertainty vanishes as the system size
is increased, i.e., the surface to volume ratio is decreased,
and this level of ambiguity is completely negligible for the modulus of
macro-scale structures.


\begin{figure*}
\includegraphics{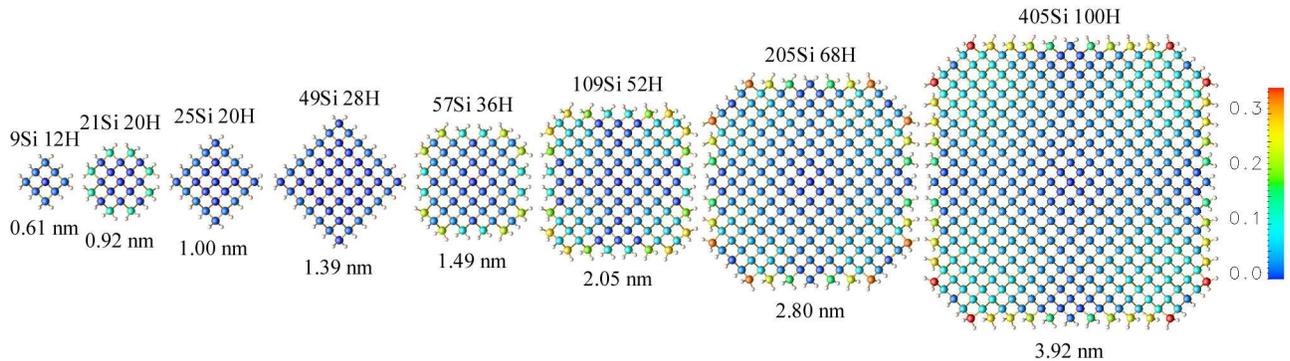}
\caption{\label{fig:wires} (Color online)
Cross sections of fully relaxed hydrogen-passivated wires,
with each Si atoms colored as shown in the legend corresponding
to its transverse relaxation in \AA.
The numbers above the wires stand for the number of atoms in the
supercell. For example, 405Si 100H means that the supercell has
405 Si atoms and 100 H atoms. The numbers below represent the wire width,
where the width is defined as the square root of the cross-sectional area.}
\end{figure*}

\section{Nanowire Calculations}
\label{sec:nwcals}

For each of the nanowire geometries shown in Fig.\ \ref{fig:wires},
the Si atoms were initially
positioned at their bulk lattice sites and hydrogen atoms were added
to terminate the bonds at the surfaces, and this configuration
was relaxed. The surface structure, such as the ground state
canted dihydride $\{$100$\}$ surface, is obtained naturally
from the relaxation and requires no initial seeding.
The supercell size of each wire is one cubic unit cell
long along the wire, and has more than 10~{\AA} vacuum space in
the transverse directions. The numbers of Si atoms and H atoms
in the supercell for each of the geometries are also
shown in Fig.\ \ref{fig:wires}.

Again the first-principles DFT code VASP has been used for
the calculations, as described above.
The energy cutoff for the plane-wave expansion is 29.34~Ry, and
1$\times$1$\times$12 Monkhorst-Pack mesh, \cite{kPointSampling}
or six points in the one dimensional irreducible Brillouin zone,
is used for $k$-point sampling.  The relaxation convergence
criterion was that the force on each atom be smaller
than $2\times10^{-3}$ eV/{\AA}
for the 1.49-nm and smaller wires, and $10^{-2}$ eV/{\AA} for the larger wires,
for which convergence of the
Young's modulus was attained with the less stringent tolerance.
The residual stress, equilibrium length and Young's modulus have
been calculated for the range of wire sizes shown in Fig.\ \ref{fig:wires}.

The extent of the range of
wire sizes is dictated by computer resource limitations.  The
calculations for the largest wire, the 3.92-nm wire, required
roughly 922 hours using typically 128 processors on the 2.3 GHz Xeon-based
MCR supercomputer, for a total of 109~285 CPU-hours for this one
wire. For the final stages of the relaxation, in which the
calculation setup as described above was applied,
it was necessary to use up to 512 processors.

\subsection{Residual stress}
\label{subsec:stress}

Residual stress can be a problem or a feature of doubly clamped
oscillators.  During etching in the device fabrication process,
stress forms in the mechanical structure.  This stress shifts
the resonant frequency of the mechanical beam, and it can affect
other properties such as dissipation.  For example, recently resonators
with very high quality
factors have been demonstrated at room temperature using
silicon nitride nanobeams under high tensile stress.\cite{Craighead2006}
Residual stress may arise due to a number of physical phenomena.  One
example is the residual stress coming from surface stress due to interactions
in the passivation layer, and it is this effect in the H-passivated
Si nanowires that we investigate now.
Within our DFT calculations, the residual stress
is equated to the axial stress when the longitudinal wire lattice
spacing is fixed at the bulk value, $\sigma _{zz} (L_0)$.

To calculate the residual stress,
the system was initially relaxed
to its zero-temperature minimum energy with the
length of the periodic supercell held fixed at the bulk lattice spacing in
the longitudinal direction. Then the relaxed total energy was
calculated for each beam in a series
of longitudinal strains at increments of roughly 0.5\%.  These total
energy values were fit to a polynomial.

We then obtain the residual stress and other important properties
from the strain derivatives of the polynomial: the first derivative
gives the axial stress, $\sigma _{zz}$,
the minimum of the polynomial, $\epsilon_{zz-eq}$,
gives the strain corresponding to the equilibrium length,
and the value of the curvature at the
minimum gives the Young's modulus $E$:
\begin{eqnarray}
\sigma _{zz} (\epsilon _{zz})  & = &  V^{-1} \partial U/\partial \epsilon _{zz}
\label{eq:axialStress}
\\
\sigma _{zz} (\epsilon_{zz-eq})  & = &  0
\label{eq:equilibLength}
\\
E & = & \left. V^{-1} \partial ^2 U/\partial \epsilon _{zz}^2\right| _{\epsilon_{zz-eq}}
\label{eq:youngModulus}
\end{eqnarray}
where $U$ is the DFT total energy.
The equilibrium length $L_{eq}$ is determined from the equilibrium strain
and the initial length $L_0$ according to
\begin{equation}
L_{eq} = L_0  \left( 1 + \epsilon_{zz-eq} \right)
\end{equation}
where here and throughout the article we are using engineering
strain, which is adequate for the small strains of interest.
The details of the equilibrium length and the Young's modulus
are discussed in the next two subsections.
We focus on the stress in this subsection.

\begin{figure}
\includegraphics[width=0.45\textwidth]{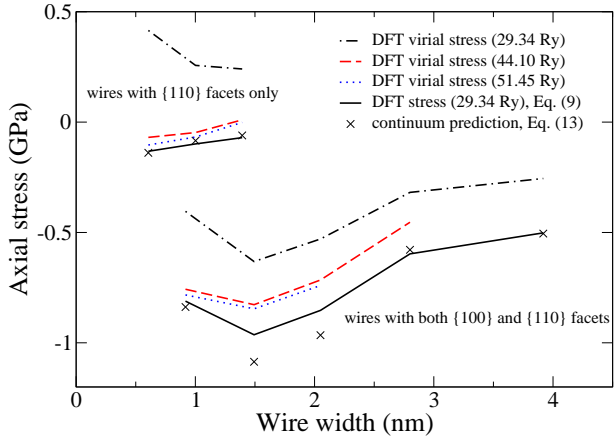}
\caption{\label{fig:stress} (Color online) Silicon nanowire axial stress as a
function of wire size calculated in DFT.
The virial stresses are obtained directly from virial formula
at the given cutoff energy indicated in the parenthesis,
whereas the DFT stress is deduced from Eq.\ \ref{eq:axialStress}.
The predictions of Eq.\ \ref{eq:resStress} are also plotted for which the
surfaces stresses are given in Table~\ref{tab:surfaceProperties}.}
\end{figure}

The size dependence of the residual stress of the H passivated
Si nanowires evident in Fig.\ \ref{fig:stress}
is driven by compressive surface stress.
The residual stress may be decomposed into
core and surface contributions.  The latter may be
further decomposed into extrinsic surface contributions from
hydrogen (H-H) interactions and intrinsic surface
contributions from the change to the Si bonds near
the surface compared to the Si bulk (Si-H and modified bond order Si-Si).
Since DFT only provides a total energy, this decomposition is
somewhat ambiguous.  Techniques such as Bond Order Potentials\cite{Pettifor}
and the Generalized Pseudopotential Theory\cite{Moriarty} have been
introduced as formalisms for extracting atomic interactions from
the underlying quantum mechanics.  They are not sufficiently developed
to apply to nanowires, however.  Local moment techniques also provide
a partition of the total energy atom by atom, but they too are
insufficiently accurate for our purposes.  Instead, we separate
different contributions to the nanowire properties through the
use of reference systems.  In the first instance, the continuum
models require core and surface properties: the uniformly strained
bulk Si systems of Section \ref{subsec:bulk} provide an approximation
to the core of the nanowire and the slab surface calculations
of Section \ref{subsec:surface} give an
approximation to the nanowire surfaces.
In a more refined analysis we decompose the surface contribution
further and estimate the extrinsic surface (H-H) interactions to be equal to
those of neighboring hydrogens in two silane molecules in the
orientation and separation of the H-passivated surface.
The intrinsic contribution is the remainder, i.e., the part
that does not come from the core or the extrinsic surface interactions.
The further decomposition of the surface effects into intrinsic and
extrinsic enables us to begin to attribute the size dependence to
specific physical processes of bonding and deformation in the
surface, sub-surface and core regions of the wire.

The extrinsic contribution to the residual stress is the most
important, as we now show.
The intrinsic surface stress is small, as expected since the dangling
Si bonds are well terminated with H atoms and the Si-Si bond order is not
significantly different than in the bulk.  The small magnitude of the
intrinsic stress is best seen in the case of the 1.39-nm
wire for which the elongation is less than 0.1\% compared to $\sim$1.5\%
of the 1.49-nm wire.  The absence of $\{$100$\}$ facets on this wire
leads to a small extrinsic stress since the H-H separation on the
$\{$110$\}$ facets is relatively large. This smallness was already
confirmed in the previous section where the surface stress of the
$\{$110$\}$ surface has been found to be negligible.
The difference between these two facets is that the vacant Si sites
above the facets are filled by one
and two H atoms on $\{$110$\}$ and $\{$100$\}$, respectively,
and the double occupancy, albeit with
$\sim$2~{\AA} H-H separation due to the shorter Si-H bond, leads to more
repulsion for $\{$100$\}$ as discussed in Sec. II.

The extrinsic surface stress due to the H-H repulsion
on the $\{$100$\}$
facets quantitatively accounts for both the compressive
residual stress $\sigma _{zz} (L_0)$
and the elongated equilibrium length $L_{eq}$ of the nanowires.
They are
related to leading order through the linear elasticity:
\begin{eqnarray}
\sigma _{zz} (L_0) & = & \sigma _{zz} (core) +
\frac{1}{A} \sum _i \tau ^{(i)}_{zz} w_i
\label{eq:resStress}
\\
\left( L_0 - L_{eq} \right) / L_{eq} & \sim & \sigma _{zz} (L_0) / E
\label{eq:eqLen}
\end{eqnarray}
where $A$ is the cross-sectional area, $w_i$ is the width,
$\tau ^{(i)}_{zz}$ is the longitudinal surface
stress of facet $i$, and $L_0$ is the bulk length of the beam.
$E$ is the Young's modulus of the beam.
In principle Eq.\ (\ref{eq:eqLen})
has anharmonic corrections, but the strains are small and the
harmonic approximation should be good.
For constant surface stress, the second term in Eq.\ (\ref{eq:resStress})
is proportional to the surface area to volume ratio; the core
stress is too, since the surface stress causes a transverse
expansion of the wire that induces a tensile core stress.
It may be seen in Fig.\ \ref{fig:wires} that surface Si
atoms on the $\{100\}$ facets undergo a substantial transverse expansion,
where for example a 0.3~{\AA} expansion amounts to
a 13\% change in the bulk Si-Si bond length. It also
induces the deformation noticeable throughout the wire, extended deep
into the core except for some of the high symmetry lines where the
expansions toward opposite directions cancel. Hence, the
transverse displacement would be hardly seen
when the $\{100\}$ facets are not present:
the $\{$110$\}$-facet wires with no $\{100\}$ facets,
compared to those with the $\{100\}$ facets,
have a negligible surface expansion and the resulting
in-plane deformation of the core.

%
%
The core stress arises as a kind of Laplace pressure and
may be estimated through a generalized Young-Laplace
equation relating the compression of the core to the surface stress.
The derivation of the expression is somewhat lengthy and
presented elsewhere.\cite{RuddLeeInPrep}
The result for the axial stress in the core averaged
across the cross section of the wire is
\begin{equation}
\sigma _{zz} (core) = - \frac{8}{\pi} \nu \tau ^{\{100\}}/ w,
\end{equation}
where $w$ is the width of the nanowire,
$\nu = C_{12}/(C_{11}+C_{12})$ is the Poisson ratio and
$C_{11}$ and $C_{12}$ are the bulk elastic constants.
Here a compressive surface stress ($\tau < 0$) leads to
to a tensile axial stress ($\sigma _{zz}>0$).

The residual stress values evaluated with a few different approaches are
plotted in Fig.\ \ref{fig:stress}. The curves denoted as `virial' are the
stresses directly obtained from the calculations via virial theorem. The
stress curve denoted simply as `DFT stress' is due to Eq.\ (\ref{eq:axialStress}),
and the one denoted as `prediction' is due to Eq.\ (\ref{eq:resStress}).
The virial stress evaluation is poor even for reasonably high cutoff
energies, i.e., the predicted stress in some cases has the opposite sign with 29.34~Ry
cutoff, and a reasonable evaluation requires a prohibitively high cutoff
energy. On the other hand, the stress obtained by the strain derivative of
the total energy with the same cutoff energy is less prone to the basis
set incompleteness critical to the direct measure, and shows better
agreement with the stress with the highest cutoff energy tested. In
addition, it is advantageous as we can obtain the stress information
of larger wires, where the cutoff energy is restricted due to computational
cost.
Considering the discrepancy between the stress due to
Eq.\ (\ref{eq:axialStress}) and the virial
stress with the highest cutoff energy, the true stress is likely to be
between the two: the convergence error for direct evaluation seems to be
negative and polynomial fitting error may be positive.

For the prediction based on the bulk and surface energies,
using the values from
Tables \ref{tab:bulkProperties} and \ref{tab:surfaceProperties}
in Eq.\ (\ref{eq:resStress})
gives predictions in very good agreement with the full nanowire
calculations as shown in Fig.\ \ref{fig:stress}.
The scatter
for 1.49- and 2.05-nm wires may be accounted for by small edge effects.
The 0.61-, 1.00- and 1.39-nm wires have no $\{$100$\}$ facets and
almost no residual stress as described above.
In the case of the second smallest (0.92-nm) wire, all of the $\{$100$\}$
atoms undergo substantial relaxation, as shown in Fig.\ \ref{fig:wires},
lowering the magnitude of the surface stress and the elongation.
This high level of agreement gives us confidence that
we understand the physics of the size dependence of the
residual stress.

\subsection{Equilibrium length}
\label{subsec:length}

\begin{figure}
\includegraphics[width=0.45\textwidth]{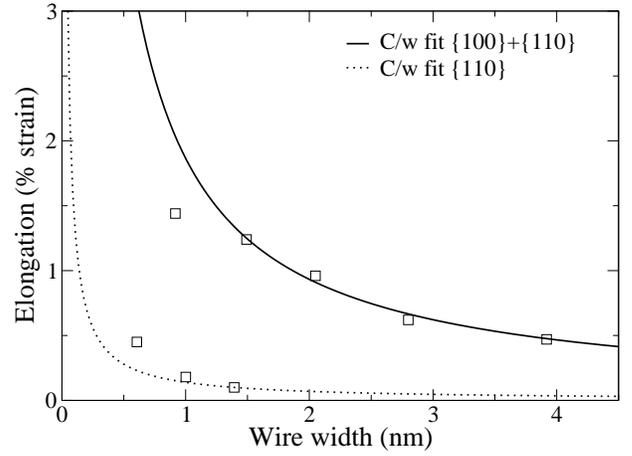}
\caption{\label{fig:elongation}Silicon nanowire
equilibrium elongation strain as a function of wire size
calculated in DFT.
The solid curve is a fit to $C/w$ of the elongation strain
to four data points from 1.49-nm and larger wires, with $C$=1.9\%-nm.
The dotted curve is a fit to $C/w$ of the elongation strain
to the 1.39-nm wire, with $C$=0.14\%-nm. }
\end{figure}

The residual stress is a direct measure of the effect of
surface stress on doubly clamped nanowires.  The analogous
property of unconstrained (free floating or cantilever) nanowires
is the equilibrium length.  It is a property that is, at least
in principle, directly measurable through x-ray diffraction.

The equilibrium elongation plotted in Fig.\ \ref{fig:elongation}
shows a systematic increase in the
elongation as the size of the wire is reduced, with the data
falling into two series.  The series are distinguished by the
amount of $\{100\}$ surface area, as discussed below.
A solid curve representing the function $C/w$ has been
superimposed on one series of data, where $w$ is the width of the wire.
The good fit of this function, with $C=1.9\%-{\mathrm{nm}}$, demonstrates
that the elongation is proportional to the surface-area-to-volume
ratio.
The axial stress shown in Fig.\ \ref{fig:stress} exhibits a similar
trend, with the stress increasing as the size of the wire decreases.
Again, the deviation of the 0.92-nm wire from the curve can be
explained by the reduction in the surface repulsion due to the
substantial relaxation around the edges. Equivalently, it can be
explained that the wire is too small to have well defined $\{100\}$ facets
that would otherwise give comparable surface effects to those on the larger wires.

As is the case with the residual stress (Fig.\ \ref{fig:stress}), the wires in
the series with elongation values closer to zero have no $\{100\}$ facets.
These wires show a significantly smaller amount of elongation compared to
those in the series with $\{100\}$ facets. The elongation for the 1.39-nm wire
is less than 0.1\%, and to achieve a comparably small amount of elongation
in a wire with $\{100\}$ facets, the wire would need to be 19 nm in
width based on the above fit.  The elongation would also be
negligible for a larger wire with the same cross-sectional shape, as
confirmed by considering that the elongation for the 2.95-nm thick
$\{110\}$ slab is less than $10^{-3}$ {\AA}.

While the elongation of the $\{110\}$-faceted wires is considerably
smaller than that of their $\{100\}$-faceted counterparts, it is actually
larger than expected: as can be seen from the dotted curve in Fig.\ \ref{fig:elongation},
the elongation of the smallest wire is even larger by roughly a factor of 2 than
the $C/w$ extension from the 1.39-nm wire, let alone any potential size-dependent effects
beyond the surface effect on the 1.39-nm wire. We do not have a conclusive explanation
for this enhanced size dependence.  It could be due to an edge
effect, but it seems more likely that it is a nonlocal surface-surface
interaction effect: for these smallest wires, each facet begins to interact
with opposite facet via electronic interactions. For example, the 1.00-nm
wire has only 5 Si layers across its cross section.  It is known from
slab calculations that the effective surface properties are modified
if the slab is too thin, and we expect an analogous effect here.

\subsection{Young's modulus}
\label{subsec:modulus}

We now consider the size dependence of the Young's modulus, the
principal subject of this work.
The Young's modulus is the second derivative of the
DFT total energy with respect to the applied longitudinal
strain divided by the volume (\ref{eq:youngModulus}).
As a second derivative, it is the most sensitive of the
quantities we compute, determining the values of the
planewave cutoff, number of $k$ points and residual force tolerance
quoted above.  Also, it is potentially susceptible to
an unwanted dependence on the polynomial fitting procedure
used for the energy.  In principle, higher order fits should
be more accurate, accommodating any anharmonicity but at the
cost of the need for additional data (more relaxed total energy
calculations).
The Young's modulus of the 1.49-nm wire
gives an indication of the sensitivity
to the order of the polynomial fit, as described in
Ref.\ \onlinecite{LeeRudd}. Fortunately, for the given order of fit,
a higher cutoff energy does not significantly improve the fit;
i.e., the largest convergence error is 0.63\% for the 2nd-order fit
compared to the highest order fit.
We find that the second order fit with 29.34 Ry energy cutoff
is reasonably good, differing by less than 2\% compared with
all the higher-order combinations tested, and it is the second
order fit that we use for the analysis of all of the nanowires,
and it permits direct comparison of the results from the entire
range of nanowire sizes up to 405 Si and 100 H atoms at a tolerable
computational cost.  We have used this technique for a systematic
study of the size dependence of the Young's modulus.

\begin{figure}
\includegraphics[width=0.45\textwidth]{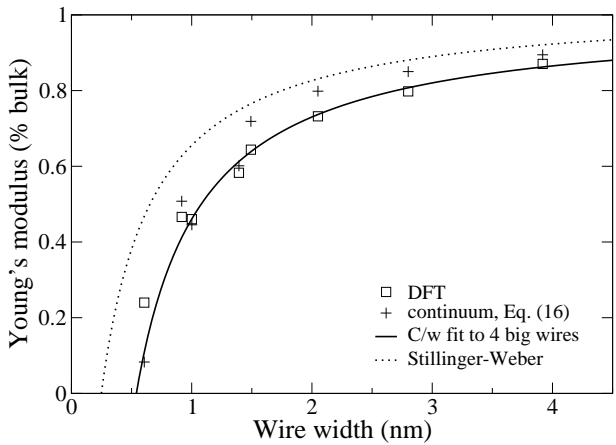}
\caption{\label{fig:modulus} Silicon nanowire Young's modulus
as a function of wire size as calculated in DFT.
For comparison the predictions of continuum formula (\ref{continuumE})
are also plotted,
for which the surface elastic constants are obtained from
the $\{100\}$ canted slabs and
the $\{110\}$ slabs. See Section \ref{sec:ref} for details.
The solid curve $E=E_{bulk}^{{\mathrm{DFT}}} - C/w$, with
$C$=66.11 GPa/nm, is a best fit to a pure surface area to volume
size dependence.}
\end{figure}

The main result is that the Young's modulus becomes softer
monotonically as the size is decreased as shown in
Fig.\ \ref{fig:modulus} and that decrease is well described
by a continuum model. It drops from the bulk value
($E_{bulk}^{{\mathrm{DFT}}}$ = 122.5~GPa)
roughly in proportion to the surface area to volume ratio.
It does not exhibit a strong dependence on the ratio of the $\{$100$\}$
to $\{$110$\}$ area seen in the equilibrium length.
The smallest Young's modulus we find for any of the nanowires
studied here is 29.4~GPa, the value for the smallest (0.61-nm) nanowire.
This value is larger than the $18\pm2$~GPa
reported from experiments on an irregularly shaped
$<10$-nm Si $\langle$100$\rangle$ nanowire.\cite{Kizuka}
Since the shapes are different, direct comparison is
challenging, but we note that the experimental stress-strain
curve is complex.  Some regions of the stress-strain curve are
in better agreement with the range of values we report here than
the region from which the value of $18$~GPa was extracted.  In any case,
they do report a softening of the Young's modulus as we have
found in the first-principles calculations.

From continuum mechanics neglecting edge and nonlocal effects,
the modulus can be expressed, slightly generalizing Ref.\ \onlinecite{Miller},
as
\begin{equation}
E = E(core) + \frac{1}{A} \sum _i S^{(i)} w_i
\label{continuumE}
\end{equation}
where $S^{(i)}$ is the surface elastic constant, a strain-derivative of
the surface stress including both extrinsic and intrinsic parts.
Thus, the Young's modulus may be decomposed into core and intrinsic, and
extrinsic surface contributions as was done for the residual stress.

We found above that the extrinsic H-H interactions dominated the surface
stress and the residual stress of the nanowires.  It is less clear
{\em a priori} what should dominate the Young's modulus, but
the fact that the modulus is insensitive to the facet ratio
(whether the $\{$100$\}$ facets are present or not)
suggests several conclusions:
\begin{itemize}
\item The core anharmonicity is irrelevant since the modulus
is not correlated with the equilibrium elongation;
\item The extrinsic contribution to the modulus
(which is strongly facet dependent)
is small;
\item The intrinsic
surface elastic constant dominates and its $\{$100$\}$ value may
be nearly sufficient to determine $E$.
\end{itemize}
Additional evidence supports each of these conclusions, as we
now show.

To quantify the core contribution,
we calculated that the Young's modulus of the bulk crystal increases
by only 1.6\% when strained
$\sim$1.5\% to match the most strained (0.92-nm) wire.
This change is negligible compared to the observed
softening (contrary to the finding that the bulk anharmonicity
dominates the size dependence of the Young's modulus of
embedded-atom-method copper nanowires \cite{bulkAnharmonicity}).

The extrinsic effect is also small, but not negligible.  Based
on silane interaction forces for the canted $\{100\}$ surface
geometry, the extrinsic contribution to the Young's modulus can be described as
\begin{equation}
\Delta E^{H-H} = \frac{1}{A} \sum _i S_{11}^{H-H} w_i^{\{100\}}
\label{extrinsicE}
\end{equation}
where $w_i^{\{100\}}$ is the width of a $\{100\}$ facet $i$.
We have estimated that the extrinsic
contribution is $\sim$8 GPa for the 1.49-nm wire,
roughly equal to E(1.49nm)-E(1.39nm), i.e., the difference
in the moduli with and without $\{100\}$ facets.
Here, the canted dihydride is particularly important in that the
initial (unrelaxed) symmetric dihydrides on the $\{100\}$ facets of
nanowires relax to a configuration very similar to the canted dihydrides
as illustrated in Fig.\ \ref{fig:wires}, only somewhat strained
due to the symmetry breaking at the edges that destroys the in-plane
periodicity of the infinite surface.
Also, given the agreement for the residual stress and
elongation between our continuum-based prediction with
the canted $\{$100$\}$ surface properties and the DFT result,
we can again confirm that the canted surface is indeed relevant
to the relaxed wires. Otherwise, the predicted residual stress and the
elongation based on the symmetric $\{$100$\}$ facet would be twice
larger than the actual value for the 1.49-nm or larger
wires.\cite{ExtrinsicComment}

\begin{figure}
\includegraphics[width=0.45\textwidth]{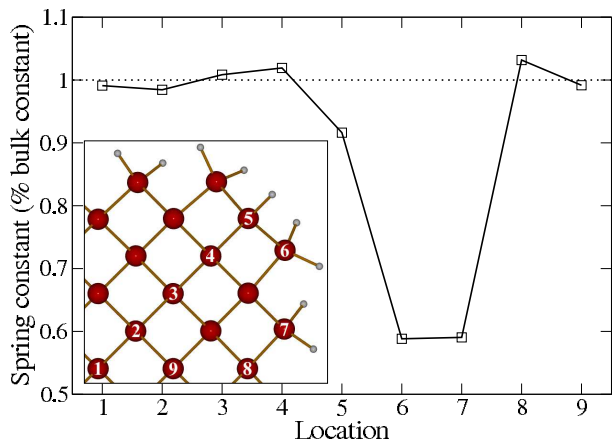}
\caption{\label{fig:springConstant} (Color online)
Spring constant of Si atoms for the 1.49-nm wire.
The inset illustrates the location of each atom in the
cross section: a quarter of the cross section is shown
with the rest of the wire related by mirror symmetry, where
atom number 1 is located at the center of the cross section.}
\end{figure}

The intrinsic contribution accounts for most of the Young's modulus.
By definition, it is the remainder once the core and extrinsic
contributions have been subtracted, and those contributions are
small as we just showed.  We may go further and create a qualitative
map of the intrinsic contribution using a bond-strength calculation akin
to an Einstein model with independent harmonic oscillators. A small
longitudinal displacement is applied to each atom, and by measuring
the induced force, the spring constant for each atom is deduced.
Each atom has three spatial degrees of freedom and hence three
oscillators, and only
the longitudinal oscillators are considered here. As can be seen
in Fig.\ \ref{fig:springConstant}, surface atoms have substantially softer
bonds. The true intrinsic effect might be even greater since
the force built up on the neighboring atoms by displacing a single
atom is relieved somewhat by the relaxation of its neighbors.
The fluctuation of the spring constant for fully coordinated atoms may
be explained by the electron density variation due to the surface relaxation
and the natural structure of silane chains, i.e., Friedel-like oscillations.
Some recent work has considered the effect of electron density variation
on mechanical properties through a bond-order approach.\cite{bondOrder}

We have also calculated
the size dependence of the modulus using Eq.\ (\ref{continuumE})
based on the surface elastic constants $S^{\{100\}}$ from the
surface reference system presented in Table~\ref{tab:surfaceProperties}.
The results, shown in Fig.\ \ref{fig:modulus}, are in good
agreement with the full first-principles calculation, and
adding the core contribution slightly improves the agreement.
The reason for the scatter for 1.49- and 2.05-nm wires is two-fold.
First, there is a variation in the tilting angle of the hydrides on
the wire surface: those around the facet center are equal or closer
to the symmetric configuration, i.e., no tilting or smaller tilting
angle.
The modulus estimation based only on the canted surface elastic
constants could overshoot the real value.
The size of the symmetric dihydride region at the middle of
the facet is determined by the usual kink analysis and is essentially
independent of the size of the facet.
Second, the scatter for 1.49- and 2.05-nm wires may be due in part
to small edge effects.
Both the effect of the symmetric dihydride region and that of the edges
would scale as the ratio of their constant areas to the facet area,
and would therefore be less important for larger wires.
Also plotted in Fig.\ \ref{fig:modulus} is the best fit curve of
Ref.\ \onlinecite{Miller} from Stillinger-Weber (SW) empirical
molecular statics calculations. The SW bulk Young's
modulus is 13\% lower and the coefficient $C$ of the $1/w$ term
is 29\% lower, representing a weaker surface effect than in 
DFT. The errors compensate for each other, 
leading to reasonable agreement for the nanoscale wires. 
This level of agreement is unexpected since the SW
potential does not have the relevant nanophysics in its
functional form or its fitting database and the strength of the
bonds does not change at the surface.  Also, the SW nanowire calculation
does not include a H-terminated surface, and thus the residual
stress and equilibrium elongation are quite different.

\section{Conclusions}
\label{sec:conclusions}

In conclusion we have found that first-principles calculation of
several mechanical properties of silicon wires predicts a size
dependence at the nanoscale.  The form of the size dependence
is in good agreement with previous models based on empirical
atomistic and continuum techniques, for all but the smallest wires
in which effects such as the electronic interaction between surfaces
are not captured in the models.  The calculations presented
here enabled an analysis of the magnitude of surface and edge
effects in the nanowire Young's modulus from first principles.
In each case the size dependence scales roughly as the
surface area to volume ratio, but for different reasons.
For the equilibrium length and residual
stress it is due to the extrinsic surface stress from interactions
in the H passivation layer; for the Young's modulus it arises from
the intrinsic contribution to the surface elastic constant.
The equilibrium length and residual stress depend strongly on
whether $\{100\}$ facets are present or not, whereas the Young's
modulus was essentially insensitive to the facet type.
Surface parameters from slab calculations capture most, but not all,
of the physics.
The size effect is not strong for the H-terminated surfaces studied here:
the Young's modulus is softened by about 50\% for a 1-nm diameter wire.
It may be possible to measure this effect directly using either
AFM deflection or resonant frequency measurements
in a double-clamped or cantilever configuration.
The change in the equilibrium length is measurable with
x-ray or electron diffraction techniques.
These effect could be substantially
stronger in silicon nanowires with different surfaces, such as bare
or oxide surfaces, making measurement easier.
These systems are more challenging for
first-principles calculation due to a greater number of candidate
structures and a greater role for charge transfer in the mechanics.
It is not clear whether the Young's modulus would increase or
decrease as the size of the beam is reduced.  There is much to
be learned still.

In the wires we have studied the surface atoms are passivated by
hydrogen atoms so that the chemical bonding is the same as in the
bulk and quantum confinement is evident,
the local electronic environment of each Si atom is uniform
throughout the wire apart from small variations.
The lack of any
noticeable quantum mechanical difference between the Si atoms in the core
and those at the surface (the Si-H interface)
makes it easier to apply continuum modeling
successfully to the hydrogen-passivated wires.
As we consider how these effects transfer to other kinds of wires,
the extrinsic surface effect
coming from the surface adsorbates will require a careful treatment,
and in cases where the surface interactions are stronger it
may prove more difficult for continuum models to provide
an accurate description of small wires.

The calculations presented here are at absolute zero temperature.
Given the present-day computers, it is not possible to carry
out these first-principles calculations at finite temperature.
Based on the results of empirical molecular
dynamics,\cite{BMVK,quartzResonators,RuddIJMSE} the general form
of the size dependence of the Young's modulus
is expected to be retained at finite
temperatures well below the melt point, but the value of the
modulus will change.  Naturally, thermal softening will shift
the entire curve, but also the value of the coefficient of
the surface-area-to-volume ratio term  will be somewhat
temperature dependent. For this and many other applications
it is desirable to construct classical interatomic potentials,
whether quantum-based or strictly empirical, that capture the
surface physics relevant to the mechanics of nanostructures.\cite{LeeSS}
First-principles calculations, such as those presented here,
lay the groundwork for the development of those potentials.
There is clearly much to be done in the development of Nanomechanics.

\begin{acknowledgments}
We would like to thank A.\ J.\ Williamson for helpful comments.
We are grateful to Livermore Computing
for extensive supercomputer resources on MCR.
This work was performed under the auspices of the U.S. Department of Energy
by the University of California, Lawrence Livermore National Laboratory,
under Contract No. W-7405-Eng-48.
\end{acknowledgments}


\begin{thebibliography}{99}
\bibitem{ExtrinsicComment}
For clarity, we emphasize that the extrinsic contribution is included
in the nanowire Young's modulus presented in this work.
\end{thebibliography}
\end{document}